# Enabling Visual Design Verification Analytics – From Prototype Visualizations to an Analytics Tool using the Unity Game Engine

Markus Borg[1], Andreas Brytting[2], Daniel Hansson[3]
[1]RISE SICS AB, Lund, Sweden, markus.borg@ri.se, +46 733 504537
[2]KTH Royal Institute of Technology, Stockholm, Sweden, andreas.brytting@gmail.com
[3]Verifyter AB, Lund, Sweden, daniel.hansson@verifyter.com

*Abstract*— The ever-increasing architectural complexity in contemporary ASIC projects turns Design Verification (DV) into a highly advanced endeavor. Pressing needs for short time-to-market has made automation a key solution in DV. However, recurring execution of large regression suites inevitably leads to challenging amounts of test results. Following the design science paradigm, we present an action research study to introduce visual analytics in a commercial ASIC project. We develop a cityscape visualization tool using the game engine Unity. Initial evaluations are promising, suggesting that the tool offers a novel approach to identify error-prone parts of the design, as well as coverage holes.

## I. Application

Design Verification (DV) is increasingly complex as advanced architectures with multi-threading multi-cores struggle with small form factor requirements. Furthermore, conflicting demands on high performance and low power exacerbate both the design complexity and its corresponding functional verification. On top of that, companies face ever-increasing pressure to decrease Time-To-Market (TTM) and the need for first pass working silicon becomes greater -- thus often DV requires most of the time in an ASIC project [9][15]. As effective and efficient DV is fundamental to maintain short TTM despite increasingly complex design, several automated approaches to dynamic DV have been developed, e.g., test suites developed according to principles from coverage-driven verification and constrained random verification.

Automated DV of complex designs, i.e., recurring execution of large regression test suites, typically leads to challenging amounts of test results. Naively automating test case execution might not bring all the expected productivity increases, as experiences from the software engineering domain reveal that large test suites require considerable maintenance [30]. Furthermore, daily execution of test suites generates an amount of test results that might be hard to overview [8], i.e., the amount of test results might exceed the DV engineer's processing capacity (aka. information overload). Traditionally, DV results are analyzed manually either directly from verification logs, test result matrices, or in simple aggregated views such as project dashboards [22]. Consequently, manual analysis tends to be either limited to a high-level, or dependent on tedious examination of textual logs.

One approach to tackle analysis of large amounts of test results is to enable visual analytics [7], defined as "the science of analytical reasoning facilitated by interactive visual interfaces" [29]. In this paper, we propose visualization of DV results using a city metaphor, i.e., presenting a cityscape, and we implement our approach in an interactive tool. Several researchers have proposed visualizing software systems as 3D cities. Wettel and Lanza developed the pioneering tool CodeCity [24], but our work is the first to visualize an ASIC project using a cityscape.

We bring visualization techniques from software engineering research to the domain of ASIC design and verification. Inspired by previous work, we develop a tool using the Unity game engine[1] to enable visual analytics of regression test data from a commercial ASIC project. Our goal is to support three main Use Cases (UC): UC1 – General exploration of large amounts of test results, UC2 – Localization of error-prone parts of the design, and UC3 – Detection of potential coverage holes.

Our work belongs in the design science paradigm, and constitutes an action research study, i.e., we aim at both improving DV at an ASIC company and studying the phenomenon from the perspective of a researcher. We previously reported from an initial evaluation of a prototype visualization [5]. In this paper, we describe how we evolved the prototype into a fully-fledged tool, designed according to Munzner's four stages of visualization design [17].

---

[1] https://unity3d.com/

The rest of the paper is organized as follows: Section II introduces visualization in general and cityscapes in particular. Section III presents the method used in our action research study. The backbone of the paper is Section IV, describing the stage-based development of our tool and our initial evaluation work. Finally, Section V concludes the paper and outlines future work.

## II. BACKGROUND AND RELATED WORK

DV is essential in any ASIC project, typically constituting the most expensive activity. Based on a large survey with 1,886 respondents in 2014, Foster reports evidence that DV on average requires 57% of the time in ASIC projects [9]. Moreover, the survey shows that average peak numbers of DV engineers outnumber design engineers in 2014 – a considerable shift in recent years, caused by increasingly challenging verification of advanced designs. In the same vein, Mehta reports that 40-50% of ASIC project resources go to functional verification [15].

A common approach to ensure high-quality DV is to define targets for both code coverage and functional coverage in the verification plan. Regression test cases that fulfill such coverage criteria are often executed as part of a daily, or nightly, build process. If modifications to a certain part of the design frequently cause regression test cases to fail, we refer to the part as *error-prone*. While low code coverage clearly suggests that a code segment has been insufficiently tested, high coverage does not necessarily mean that its corresponding test suites are particularly effective at exposing defects [13] – defects might slip through even though the design code was exercised. Consequently, DV engineers need a way to analyze whether there appears to be any deficiencies in the test system, i.e., localizing potential *coverage holes* where defects might slip through the DV activities. The rest of this section introduces the previous work that we rely on to support visual analytics in an ASIC project.

### A. *Visualization Fundamentals*

Visualization includes any technique for creating images or diagrams to communicate a message. Creating successful representation of data has been a research target for decades. Tufte, one of the most cited visualization scholars, has formulated several principles and guidelines [23]. Tufte's advice include: 1) focus on the data: above all else, show the data, i.e., "maximize the data-ink" ratio, do not visualize things that are not present in the data and 2) avoid "chartjunk", i.e., visual elements that are not necessary to comprehend the data. Adhering to Tufte's recommendations often mean avoiding 3D and artistic color schemes, but in this paper, we present an example in which we believe a visual representation in colorful 3D indeed is useful. The rationale is that humans are familiar with 3D in the physical world, thus the extra dimension can help perceiving complex information.

Software visualization means using visualization to show either static perspectives of software systems or their dynamic run-time behavior. Diehl divides this into three categories [6]: 1) *structure* refers to static parts and relations of the software system, i.e., the source code modules and the static call graph, 2) *behavior* refers to the execution of the software, i.e., memory allocation or communication between objects in object-oriented languages, and 3) *evolution* refers to the development process leading to a software system, i.e., the changes to the source code or the results from testing during development. The work we present in this paper is an example of software visualization belonging to the evolution category. We propose to bring visualization techniques proposed in software engineering contexts to the domain of DV.

### B. *Cityscapes in Software Visualization*

Wettel and Lanza pioneered using cityscapes in static software visualization [24]. They developed the tool CodeCity to enable interactive analysis of object-oriented software as 3D cities. In CodeCity, classes are represented as buildings in the city, while the packages are presented as districts. Each building has a height mapped to the number of methods in the corresponding class and the base size shows the number of attributes. In the tool, a number of different source code metrics can be visualized using colors of the buildings, e.g., the lines of code. The authors conducted a controlled experiment [26] with 10 individual tasks related to software maintenance and report that their subjects completed them better (+24% correctness) and faster (-12% completion time) compared to a control group working with Eclipse and Excel. However, the improvement was only seen on tasks that required a big picture overview of the system under study, not on focused tasks based on detailed information. Furthermore, CodeCity does not provide support for the three UCs targeted in this study.

Several other authors have developed approaches to visualize software as cityscapes. Biaggi developed Citylyzer, a Java port of CodeCity as an Eclipse plugin [4]. A third example of an Eclipse plugin for source code visualization is Manhattan, developed by Bacchelli *et al.* [1]. Their cityscapes focus on showing metrics related to product quality and development site productivity. Balogh and Beszedes developed CodeMetropolis, a cityscape that is created using the computer game Minecraft [2]. The authors decided to use Minecraft for visualizing the cityscape because of its "high quality graphics and expressive power" combined with its support for third party software integration. Merino *et al.* also used a game engine to create cityscapes [16]. Their tool CityVR uses Unity to create an immersive interactive experience using virtual reality.

Some authors have also used cityscapes to visualize test results. Balogh *et al.* proposed using their tool CodeMetropolis to visualize test-related metrics [3], e.g., code coverage and advanced test metrics such as partition, specialization, and uniqueness of test cases. Sosnowka developed another approach to visualize the system under test, focusing on low level test cases [20]. He presents the number of test cases, execution status, modification dates, and number of executions on the cityscape.

The publications by Merino *et al.* and Balogh *et al.* constitute the most similar previous work. Both publications use game engines to create cityscapes for software visualization – CityVR is even developed in Unity. Balogh *et al.*'s work, on the other hand, uses a game engine to visualize aspects related to software testing. Consequently, our work combines aspects from both publications into a tool, created by Unity, enabling visual analytics of regression test results. Moreover, in contrast to previous work, we focus our work on DV in ASIC projects.

### III. METHOD

Our work belongs to the *design science research paradigm*, i.e., we develop new and innovative artifacts to understand and improve DV engineers' interaction with test results. Design science is an established approach in information systems research and management research, but the ideas have been adopted also in engineering research [21]. Hevner *et al.* expresses that the main purpose of design science research is to achieve knowledge of a problem domain by building and application of a designed artifact [11]. Van Aken stresses the solution-oriented aspects by stating that the main goal of design science research is to "develop knowledge that the professionals of the discipline in question can use to design solutions for their field problems" [24]. A central component in any design science research is the build-evaluate loop, i.e., several iterations are used to reach the final artifact.

While design science captures the gist of our work, the actual research method that guided our study was *action research*. In action research, originating in the social sciences, the objective is not to objectively study a real-world phenomenon, but to influence it and learn from the process [19]. Action research stresses an iterative research endeavor with dual goals, i.e., the goals are both to solve a real-world problem and to study it from the perspective of a researcher. Other fundamental aspects of action research include: 1) the researcher-client agreement, i.e., the explicit ambition to both improve a specific real-work situation and to advance research, 2) the cyclical project execution, i.e., iterations encompassing planning, taking action, and evaluation, 3) connection to theoretical frameworks, i.e., the body of related research, and 4) learning through critical reflection.

Our work adheres to Wieringa's adaptation of action research for the engineering domain, referred to as *technical action research* (TAR) [27]. According to Wieringa, the key to successful TAR is to keep three roles separate during the work. The roles are: 1) *the designer* – developing the novel artifact, 2) *the helper* – using the artifact to support the client, and 3) the researcher – documenting lessons learned from the process. In our work, the three authors of the paper embody the separation. The first author is the researcher, who designed the study and is now responsible for reporting the results in this paper. The second author is the designer, who developed the visualization tool. Finally, the third author is the helper, who communicated directly with the client and acted as a client proxy between meetings.

Figure 1 illustrates our research method. The three co-authors, represent the three roles (designer, helper, and researcher), collaborating in the center of the build-evaluate loop. A cycle of the build-evaluate loop typically lasted one week. The tool development was motivated by both 1) goals originating in the environment, i.e., the need for better analysis of regression test results in the client's ASIC project, and 2) theory from the knowledge base, i.e., related work on ASIC design, regression testing, and visual analytics. We provided our evolving visualization tool to the ASIC project (however, often using a client proxy) and now contribute lessons learned to the knowledge base through this paper.

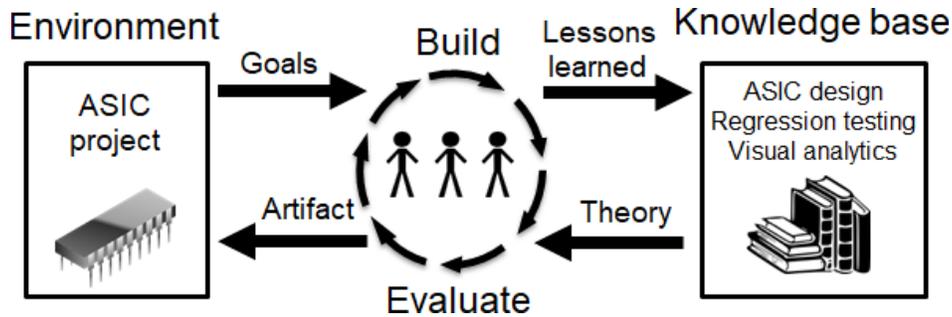
Figure 1: Overview of the research method.

The tool development in the build-evaluate loop was organized in four Design Stages (DS) in line with Munzner's recommendations for visualization design [17]. The first stage is to characterize the problem and the available data (DS1: Problem Characterization). The author with the role as helper acted as a knowledgeable client proxy and also knew the available data – regression test results from the tool PinDown [10]. The second stage involves mapping the problem characterization to an information visualization problem (DS2: Mapping to Solution Concept). During this stage, we created prototype visualizations for localization of error-prone parts and coverage hole detection. Two of the client's senior DV engineers validated the prototype visualizations [5], thus we considered the prototypes successful proofs-of-concept motiving continued research. According to Munzner, the third stage is the design of visual encoding and interaction (DS3: Solution Design) and the fourth stage covers the implementation of the solution design with an effective algorithm (DS4: Design Implementation). In this paper, we describe how we accomplish the final two stages by developing a tool using the Unity game engine.

## IV. RESULTS

This section describes the development of the tool, organized according to Munzner's recommendations for visualization design [17].

### A. DS1: Problem Characterization

The case project under study is an ASIC project at a major manufacturer. Table 1 shows descriptive statistics about the project. We study six months of commit history and its corresponding results from regression testing. Ninety-six engineers committed changes to the code base, encompassing 61,200 files in total – dominated by Verilog and SystemVerilog. The regression test suite contains 500 test cases and is executed 3-8 times per day, i.e., roughly 1,500-4,000 results from individual test cases are generated every day.

Traditionally, results from regression testing are aggregated in a test result matrix. Individual cells in the matrix show the test result from executing a test case (or test suite). A DV engineer can examine detailed execution results from individual test cases, e.g., by investigating test logs, but the test result matrix does not allow any further analysis. However, regression testing over the duration of a project might reveal valuable patterns -- if presented properly. The DV engineers in the case company struggle with two primary goals: 1) directing design engineers to parts of the source code that frequently fail in the DV, indicating low quality, and 2) identifying potential coverage holes that should be targeted by additional regression test cases.

Table 1: Descriptive statistics of the ASIC project.

|  |  |  |
|---|---|---|
| Code base | programming languages | Verilog, SystemVerilog, assembler, C, shell scripts |
|  | #files | 61,200 (3.8 GB) |
|  | commit history | Oct 2016 – Mar 2017 |
|  | #commit sets | 1,544 |
|  | #committers | 95 |
| Regression tests | #test cases | 500 (Constr. Random Test) |
|  | execution time | 2 hours |
|  | execution frequency | 3-8 times/day |

**Available data**. The collected data contain 1) the ASIC project's entire *commit history* from the version control system and 2) *DV results* from automated regression testing. The data were extracted from Perforce and the debug tool PinDown [10], respectively. The commit history covers 1,544 commit sets containing one or several modified files. Each commit set has a unique identifier, a time stamp, and a list of modified files. The DV results contain the identifiers of 68 commit sets that triggered a failed regression test, complemented by their commit messages. We refer to these 68 commit sets as "bad commits". By combining the commit history with the DV results, we created a list of modified files that were part of bad commits, i.e., "bad files". Note that a bad file did not necessarily cause a test case to fail, but it had been part of a bad commit.

The data that we visualize contain 7,099 modified files out of which 2,537 are bad files. Each file has a full path file name. We use unique modified files as the most fine-granular item in our data, and calculate the fraction of commits for which the file had been part of bad commits (in contrast to being a part of a commit set that did not cause a regression test case to fail). We refer to this fraction as the "badness ratio", ranging from 0%-100%.

Coverage data is not available in this data set. Code coverage requires the code to be instrumented, something which is not done for many regression test suites, where the goal is just to detect test failures. Functional coverage is available only if the test suite consists of at least some constrained random tests. In this paper, we instead use a way to analyze coverage without having access to actual coverage data: we look for folders with suspiciously low "badness ratio" compared to the number of file commits. This allows us to detect functional coverage holes where the engineer has missed specifying some important functional coverage points. A functional coverage hole in this paper is consequently not limited to finding specified coverage points that have not been covered by any test cases.

*B. DS2: Mapping to Solution Concept*

Figure 2 shows our prototype visualization. We use Unity to create a cityscape as a unique view of the design under test. Each building shows a file that has been part of a commit set at least once during the project. The buildings are structured according to their file paths, i.e., the city blocks represent the project's folder structure. The hierarchical folder structure is presented as elevated ground in the city, see a) in Figure 2. The layout of city blocks minimizes the space required to fill a rectangular area, i.e., it does not correspond to the design (or physical) hierarchy of the ASIC, although this is a promising direction for future work.

The backbone of the visual analytics enabled by the visualization is twofold. First, the higher the buildings, the more often the corresponding file has been part of commit sets. Due to the skewed distribution of commits per file, we use a logarithmic scale for the height of buildings. Second, we use the red-yellow-green combination to illustrate the badness ratio, as it has previously been shown intuitive when visualizing software testing [14].

**Initial validation**. Based on the prototype visualization, we conducted an initial validation of our general visualization approach – constituting a sanity check of our cityscape view of the design under test. Two senior DV engineers at the client, both actively working with the design, explored the prototype in an informal one-hour meeting guided by four open-ended questions (Q). We discussed the general overview, as well as error prone areas and potential coverage holes. The results from the validation has been reported in a previous publication [5], but to keep this paper self-contained, we summarize the findings below.

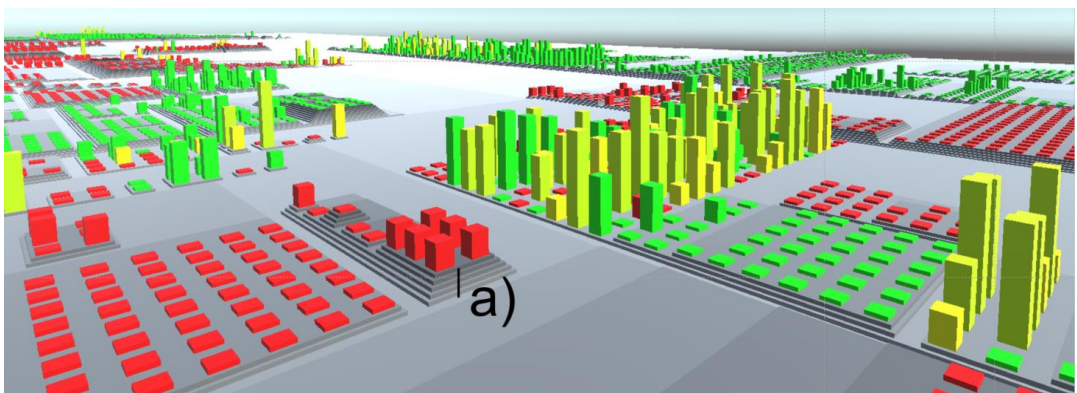

Figure 2: Prototype visualization.

**Q1. Are the potential coverage holes valid?** Yes, the prototype highlighted parts that might not have been tested enough. However, the largest potential coverage hole corresponded to a test bench under development -- it was not supposed to be tested in this specific project.

**Q2. Are the error-prone areas valid?** Yes. The engineers were surprised that the most error prone folders were not in the RTL (Register Transfer Level) code, containing the most code churn [18], but instead in a specific part of the test bench. This finding triggered initial analysis on the client side.

**Q3. Is the cityscape a useful visualization?** Yes. The engineers would like to conduct visual analytics using cityscapes on a recurring basis, e.g., analyzing each ASIC project every 6 months. Also, a cityscape could be part of projects' post-mortem analyses.

**Q4. Any improvements suggestions?** Yes. First, the user must be able to filter individual commit sets such as large software merges, i.e., commit sets that encompass very many individual files. Software merges are critical operations that often cause at least some regression test cases to fail. The evaluation highlighted the importance of intuitive interaction, in line with conclusions by Engström *et al*. [7], i.e., users should be able to filter results and focus attention accordingly. Second, the two engineers requested intuitive navigation and an option to export a textual representation of the visualization.

*C. DS3 and DS4: Solution Design and Implementation*

We evolved the prototype visualization into a fully functional visualization tool using Unity. The motivation for using a game engine, and in particular Unity, is threefold. First, interaction is fundamental for successful visual analytics [7] – and interaction design is a primary purpose in game development. Second, contemporary game engines scale to visualize very large realistic worlds, i.e., very large quantities of historical test results can be presented. Third, Unity is an established solution that powers games such as Cities: Skylines and Pokémon GO, but still it offers a user-friendly game development platform with support for C# and JavaScript -- and a very active online community.

Figure 3 shows a screenshot of the user interface and a high-level visualization of the design under test. The tool offers two types of information filtering: 1) setting a specific time window for commits (cf. a) in Fig. 3) and 2) hiding an entire commit set (an item in a context menu appears when clicking a building, not in Fig. 3). Information filtering is critical to let the user actively explore the data. Note that we do not restructure the physical layout of blocks in the cityscape, we only remove (or reduce the height) of buildings.

The main purpose of the tool is to help DV engineers identify error-prone parts and potential coverage holes, both aiming for a folder-level granularity. The main features supporting this task are the highlight windows in the upper right corner (cf. b) in Fig. 3. First, to find error-prone parts, the user can set thresholds values for the minimum number of commits and the minimum badness ratio. When the highlight feature is toggled, the lowest level folder fulfilling the criteria is illuminated by a red search light (cf. c) in Fig. 3). Second, we developed an analogous feature for potential coverage holes, i.e., the user can set threshold values for the minimum number of commits and the maximum fault rate. When the corresponding highlighted is toggled, yellow search lights signal the matching folders in the design (cf. d) in Fig. 3).

User navigation in the cityscape relies on an established control scheme from 3D games: WASD controls with mouselook. The mouse is used to direct the camera and the four WASD keys are mapped to movement: W – move forward, S – move backward, A – move left (without turning), and D – move right (without turning). In addition, we implemented height navigation using the up and down arrow keys. The user can freely fly around the city to closely examine buildings and neighborhoods to better understand the status of the design under test. We hypothesize that implementing a control scheme commonly used by the gaming industry should lower the entry bar for at least a subset of DV engineers.

The user can directly interact with the cityscape using the mouse. When hovering the mouse cursor over a building or a city block, a tooltip appears with summary information (cf. e) in Fig. 3), e.g., the total number of commits and a list of bad commits. When clicking on a building, a context menu appears enabling information hiding for all involved commit sets (as described earlier).

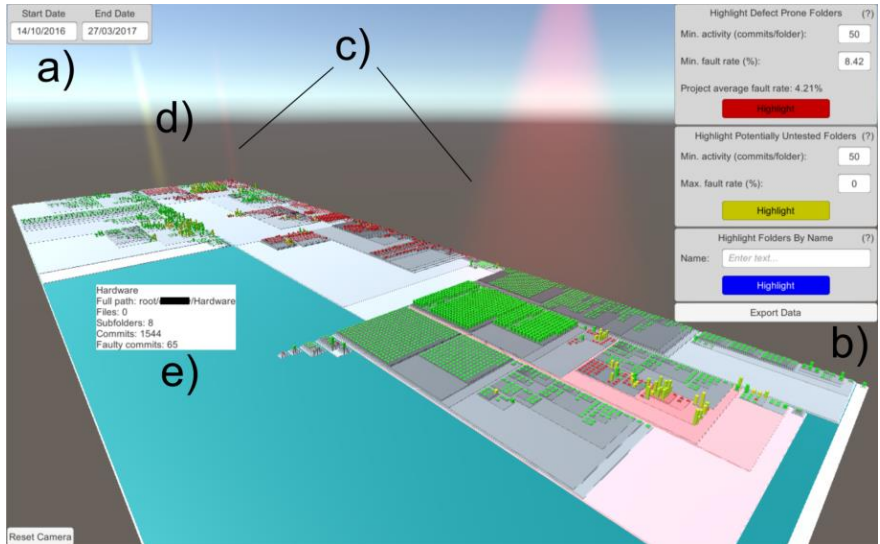

Figure 3: User interface of the visualization tool and an overview of the design under test.

**Evaluation protocol**. In line with common practice in evidence-based medicine [12], we report the evaluation protocol prior to conducting the evaluative study. This practice enables peer-review of its design, prior to running effort-intensive studies with human subjects. We plan to evaluate our tool using a combination of qualitative and quantitative research methods, i.e., in-depth interview sessions with DV engineers and a controlled experiment with student subjects.

The in-depth interviews will resemble our initial user evaluation. A handful of local DV engineers, all with considerable experience, have agreed to take part in our evaluation. We plan to invite them to individual evaluation sessions, introduce them to the research we are conducting, and provide them with short tool instructions. We will then study how the DV engineers interact with the tool as they complete three tasks: 1) exploratory navigation of the design under test, 2) suggesting parts that should be further analyzed by design engineers for fault proneness, and 3) suggesting parts that should be targeted by DV engineers for additional testing. The study will be conducted in a controlled setting and we will follow a think aloud protocol, a standard method in usability testing, and record all audio and capture the screen during the sessions. The collected data will primarily be analyzed qualitatively using thematic analysis [28].

The controlled experiment with student subjects will be conducted at Lund University, Sweden, mainly targeting students in MSc students in electrical engineering. We aim to organize multiple sessions in class-room settings with roughly 25 subjects that will be randomized into two separate single-blind treatment groups: 1) the visualization group (supported by our tool), and 2) the control group working with traditional test result matrices. The subjects will complete tasks related to design and DV management, i.e., allocate limited resources (analogous to "distribute $100" tasks) to improve the current level of testing and to assign design engineers to investigate specific parts of the code. We will measure the time required to complete the task, and the folders selected by the visualization group and the control group, respectively.

## V. CONCLUSIONS

Pressing market demands combined with advances in semiconductor technology has greatly increased the complexity of modern ASIC architectures. For Design Verification (DV) to keep up with the increased complexity, automation is a key solution, e.g., build systems and regression testing. However, increasing the number of regression test executions puts additional pressure on the DV engineers analyzing the test results. On the other hand, successful overviews of regression testing could enable new levels of analyzability.

We present a tool, developed in the Unity game engine, that enables visual analytics in DV. The tool supports exploratory navigation of the design under test, as well as identification of error-prone parts of the design and potential coverage holes. The initial evaluation of a visualization prototype based on six months of data from a commercial ASIC project was successful. Based on the user feedback, we posit that two of the well-known challenges of 3D visualizations, i.e., interaction and occlusion (information hiding), are mitigated by using a modern game engine. Our next step is to validate this claim – in this paper we present a proposed tool evaluation protocol.

We argue that visual analytics in DV is a promising direction for future research. Well-designed visualizations could help DV engineers to stay on top of regression test results in large projects and help managers focus available design and verification resources. Furthermore, using a game engine simplifies the interaction design needed for successful visual analytics. Finally, we argue that ASIC projects might be better suited for cityscape visualizations than software projects, as ASICs actually have a physical design. Future versions of the tool should explore using the design or physical hierarchies to structure the blocks in the cityscape.

ACKNOWLEDGEMENTS

ACKNOWLEDGEMENTS

Thanks go to our client for supporting our work, in particular the two ASIC DV engineers involved in the initial user study. This work has been financially supported by the ITEA3 initiative TESTOMAT Project through Vinnova.